\documentclass{article}
\usepackage{hyperref} 
\usepackage[english]{babel}
\usepackage{amsmath}    
\usepackage{subfigure}
\usepackage{graphicx}
\usepackage{tabularx}
\usepackage{booktabs}
\usepackage{array} 
\usepackage{caption} 
\usepackage{authblk}  
\usepackage[letterpaper,top=2cm,bottom=2cm,left=3cm,right=3cm,marginparwidth=1.75cm]{geometry}

\usepackage{amsmath}
\usepackage{graphicx}
\title{Extreme Value Statistics of Community Detection in Complex Networks with Reduced Network Extremal Ensemble Learning (RenEEL)}
\author[1,2]{Tania Ghosh}
\author[1,3]{R.K.P. Zia}
\author[1,2,4]{Kevin E. Bassler*}

\affil[1]{Department of Physics, University of Houston, Houston, TX 77204}
\affil[2]{Texas Center for Superconductivity, University of Houston, Houston, TX 77204}
\affil[3]{Center for Soft Matter and Biological Physics, Department of Physics, Virginia Tech, Blacksburg, VA 24061}
\affil[4]{Department of Mathematics, University of Houston, Houston, TX 77204}

\begin{document}

\maketitle
\noindent *Correspondence: bassler@uh.edu; Tel.: +01-713-743-3568 (K.E.B.)
\begin{abstract}
Arguably, the most fundamental problem in Network Science is finding structure within a complex network. One approach is to partition the nodes into communities that are more densely connected than one expects in a random network. ``The” community structure then corresponds to the partition that maximizes Modularity, an objective function that quantifies this idea. Finding the maximizing partition, however, is a computationally difficult, NP-Complete problem. We explore using a recently introduced machine-learning algorithmic scheme to find the structure of benchmark networks. The scheme, known as RenEEL, creates an ensemble of $K$ partitions and updates the ensemble by replacing its worst member with the best of $L$ partitions found by analyzing a simplified network. The updating continues until consensus is achieved within the ensemble. We perform an empirical study of three real-world networks to explore how the Modularity of the consensus partition depends on the values of $K$ and $L$ and relate the results to the extreme value statistics of record-breaking. We find that increasing $K$ is generally more effective than increasing $L$ for finding the best partition.
\end{abstract}
\vspace{10pt}

Finding the community structure within a complex network that relates to its function or behavior is a fundamental challenge in Network Science~\cite{Fortunato_2010, Newman_2004_A}.
It is a highly nontrivial problem, as even defining what one means by "the structure" must be specified~\cite{Schaub_2017, Peel_2017, Wolpert_2005}. A variety of approaches can be used to partition the nodes of a network into communities. Each approach will divide the nodes according to a different definition of structure and will, in general, find a different partition. Which approach finds the correct or best structure depends on the particular network problem being considered. 
One commonly used approach is to partition the nodes into communities that are more densely connected than expected in a random network. 
In this approach, the community structure corresponds to the partition that maximizes an objective function called \textit{Modularity}~\cite{Newman_2004_A,Newman_2006_B}. For a given nodal partition $C\equiv \{c_1,c_2,\ldots \}$, Modularity $q$ is defined as
\begin{equation}
    q(C) = \frac{1}{2m} \sum_{\langle ij \rangle} \Bigg[A_{ij}-\frac{k_ik_j}{2m}\Bigg]\delta_{c_ic_j},
    \label{eq:1}
\end{equation}
where the sum is over all pairs of nodes $\langle ij \rangle$, $c_i$ is the community of the $i$th node, and $m$ is the total number of links present in the network. $k_{i}$ and $A_{ij}$ are respectively the degree of $i$th node and the 
$ij$th element of the adjacency matrix. 
Thus, Modularity is the difference between the fraction of links inside the partition's
communities, the first term in Eq.~\ref{eq:1}, and what the expected fraction would be if all links of the network were randomly placed, the second term in Eq.~\ref{eq:1}.
The task is to find the partition $C$ that maximizes $q$. We denote the maximum value of $q$ as $Q$, the value of which is called ``the Modularity" of the network. Community detection by Modularity maximization is both intuitively appealing and has been justified by comparing its results with prior ``ground-truth" knowledge of structure in a variety of networks~\cite{KuninJN_2023, ZAMANIESFAHLANI2021}.

However, finding the partition that maximizes Modularity is a computationally difficult NP-complete problem~\cite{Brandes_2008,Brandes_2006}. 
Finding a guaranteed exact solution for a large network is, thus, generally infeasible. It is, therefore, important to develop an approximate algorithm that has polynomial time complexity, i.e., is fast, and that finds near exact best solutions for large networks, i.e., is accurate. A number of such algorithms have been developed. 
They range from very fast but not so accurate algorithms, such as the Louvain method~\cite{Blondel_2008} or randomized greedy agglomerative hierarchical clustering~\cite{Ovelgnne_2010}, to more accurate but slower algorithms~\cite{Sun2009}, such as one that combines both agglomeration and division steps~\cite{Trevino_2015,Liu_2019}. The accuracy of all of the algorithms tends to decrease as the size of the network increases.

All of the fast Modularity maximizing algorithms are stochastic because at intermediate steps of their execution there are seemingly equally good choices to make that are randomly made. In the end, those choices can be consequential because different runs of an algorithm with different sets of random intermediate choices can result in different solutions. Because of this, multiple runs of an algorithm are often made, say 100, to analyze a network, producing an ensemble of approximate partitions. The partition in the ensemble with the largest Modularity is then taken as the network's community structure, while all other partitions in the ensemble are discarded. 

Recently, an algorithmic scheme has been introduced that uses information in an ensemble of partitions to produce a better, more accurate partition when seeking to maximize an objective function, such as Modularity. The scheme,  Reduced Network Extremal Ensemble Learning (RenEEL)~\cite{Guo2019}, 
uses a machine-learning paradigm for graph partitioning, Extremal Ensemble Learning (EEL). 
EEL begins with an ensemble $\mathcal{P}_K$ of $k \leq K$ unique partitions. 
It then iteratively updates the ensemble using extremal criteria until consensus is reached within the ensemble about what the ``best" partition is, e.g., the one with the largest Modularity.
Each update considers adding a new partition $P$ to $\mathcal{P}_K$ as follows.  
If $P\in \mathcal{P}_K$, then the ``worst" partition in $\mathcal{P}_K$, $P_{\rm worst}$, is removed from $\mathcal{P}_K$, reducing the size of $\mathcal{P}_K$ by one $k\rightarrow k-1$, and the update is complete.
If $P\notin \mathcal{P}_K$ and $P$ is worse than $P_{\rm worst}$, then again $P_{\rm worst}$ is removed from $\mathcal{P}_K$, reducing the size of $\mathcal{P}_K$ by one $k\rightarrow k-1$, and the update is complete.
If $P\notin \mathcal{P}_K$ and $P$ is better than $P_{\rm worst}$, then 
if $k=K$ $P_{\rm worst}$ is replaced by $P$ in $\mathcal{P}_K$ and the update is complete, or if $k<K$ $P$ is added to $\mathcal{P}_K$, increasing the size of $\mathcal{P}_K$ by one $k\rightarrow k+1$, and the update is complete.
Iterative updates continue until $k=1$. The remaining partition is the consensus choice for the best partition.

To ensure fast convergence to a consensus choice in the EEL updates, RenEEL conserves the consensus within $\mathcal{P}_K$ that exists up to that point each time it finds a new partition to be used in an update. It does so by partitioning a \emph{reduced network} rather than the original network. The reduced network is constructed by collapsing the nodes that every partition in $\mathcal{P}_K$ agrees should be in the same community into ``super" nodes. Reduced networks are smaller than the original network and can be analyzed faster, spending effort only to decide how to improve the partitioning where there is disagreement within $\mathcal{P}_K$. The consensus within $\mathcal{P}_K$ increases monotonically, and the size of the reduced networks decreases monotonically as the EEL updates are made. RenEEL creates an ensemble $\mathcal{P}_L$ consisting of $L$ partitions found by analyzing the reduced network to decide the partition to use in each EEL update. The best partition in $\mathcal{P}_L$ is then used in the update.

There is wide flexibility within the RenEEL scheme. A \emph{base algorithm} is used to find the partitions that initially form the $\mathcal{P}_K$ ensemble and those that form each $\mathcal{P}_L$ ensemble. The base algorithm can be any Modularity-maximizing algorithm. Multiple base algorithms can even be used.   
There is also freedom to choose the values of $K$ and $L$, respectively, the maximum size of $\mathcal{P}_K$ and the size of each $\mathcal{P}_L$. The best base algorithm choice and values of $K$ and $L$ depend on the network being analyzed, desired accuracy, and available computational resources.

This paper investigates the effect of varying $K$ and $L$. Larger values of $K$ and $L$ will typically lead to a final, consensus best partition with larger values of Modularity $Q$~\cite{Guo2019}. But how does the value of $Q$ of the consensus partition typically depend on $K$ and $L$? How quickly is the value of $Q$ found expected to approach $Q_{\max}$, the Modularity of the exact best partition? Given only limited computational resources, is it better to increase $K$ or $L$? We empirically study the community structure of three well-known real-world networks: Networks A, B, and C, to answer these questions. 

Network A is the As-22july06 Network~\cite{Newman_2011}. It is a snapshot in time of the structure of the Internet at the level of autonomous systems. It has 22,963 nodes which represent autonomous systems and 48,436 links of data connection. Network B is the PGP Network~\cite{Newman_2011}. It is a snapshot in time of the giant component of the Pretty-Good-Privacy (PGP) algorithm user network. It has 10,680 nodes which are the users of the PGP algorithm and 24,316 links as the interaction among them. Lastly, Network C is the Astro-ph network. It is a coauthorship network of scientists in Astrophysics consisting of 16,706 nodes representing scientists and 12,1251 links representing coauthorship in preprints on the Astrophysics arXiv database~\cite{Newman_2001}. 
\begin{figure}
\centering
\includegraphics[width=10.5 cm]{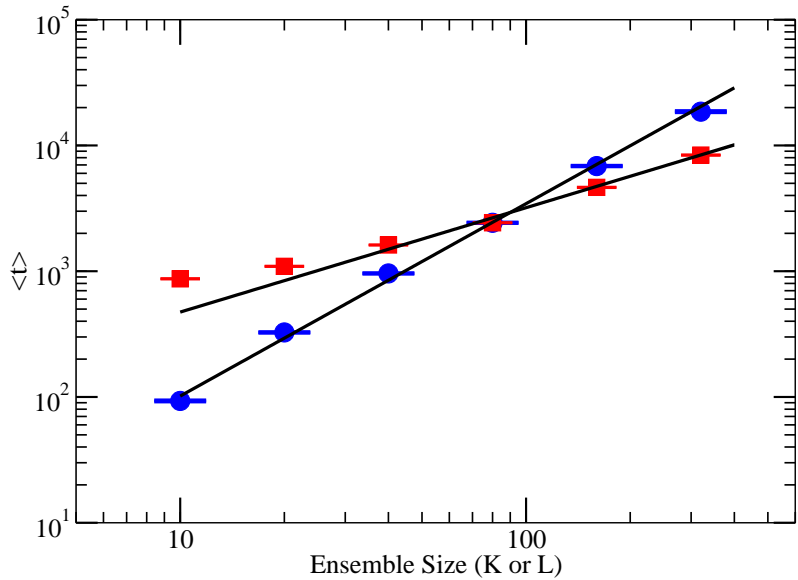}
\caption{ The compute time for RenEEL to complete when analyzing Network A. Data are measured in seconds. Blue circles are results as a function of $K$ for fixed $L=80$, and red squares are results as a function of $L$ for fixed $K=80$. The straight lines are power-law fits to the data of ensemble sizes greater than 10. The slope of the fit to the blue circles is $\alpha_K=1.489(4)$, and to the red squares is $\alpha_L=0.810(3)$. }
\label{time}
\end{figure}
\unskip
\vspace{3mm}
The RenEEL algorithm we use to study these networks employs randomized greedy agglomerative hierarchical clustering~\cite{Ovelgnne_2010} as the base algorithm. For the three networks, 300 different runs of RenEEL were made for independent values of $K$ and $L$ of 10, 20, 40, 80, 160, and 320, respectively. The compute time required to find the consensus partition was measured for each run.
The mean and standard error of the compute times for the runs at a given value of $K$ and $L$ were then calculated. The full results are listed in tables~\ref{tab:Apen1}, \ref{tab:Apen2}, and \ref{tab:Apen3} in the Appendix. For a fixed value of $L$ or $K$, we find that the mean compute times $\langle t \rangle$ increase asymptotically as a power of the other ensemble size, 
\begin{equation}
    \langle t \rangle \left. \sim K^{\alpha_K} \right|_{L\;{\rm fixed}}
    \quad
    \mbox{and}
    \quad
        \langle t \rangle \left. \sim L^{\alpha_L} \right|_{K\;{\rm fixed}}.
        \label{eq2}
\end{equation}
For example. Fig.~1 shows this power-law behavior for Network A when $L$ and $K$ have fixed values of 80.
Two-parameter, nonlinear least squares fits to data for ensemble sizes greater than 10 were then used to determine the proportionality constant and $\alpha$. 
Tables~\ref{tab:table1} and \ref{tab:table2} show the values for $\alpha_K$ and $\alpha_L$, respectively, that result from fits at different fixed values of $L$ and $K$ for each of the three networks.
All statistical errors reported in this paper are $\pm 2\sigma$.

\setcounter{table}{0}
\renewcommand{\thetable}{I\alph{table}}
\begin{table}
\captionsetup{justification=raggedright, singlelinecheck=false} 
\caption{\label{tab:table1}
Values of scaling exponent $\alpha_K$ of the compute time divergence at fixed values of $L$.}
\newcolumntype{C}{>{\centering\arraybackslash}X}
\begin{tabularx}{\textwidth}{CCCC}
\toprule
\textbf{$L$} & \textbf{Network A}	& \textbf{Network B}	& \textbf{Network C}\\
\midrule
20 &      1.382(2) &        1.262(2) &         1.331(4)\\
40 &      1.431(2)  &       1.212(2) &         1.300(4) \\
80 &      1.489(4) &       1.240(2) &         1.300(2) \\
160 &     1.483(2) &       1.164(2) &         1.320(4) \\
320   &   1.469(2) &       1.194(4) &         1.311(6) \\
\bottomrule
\end{tabularx}
\end{table}

\begin{table}
\captionsetup{justification=raggedright, singlelinecheck=false} 
\caption{\label{tab:table2}
Values of scaling exponent $\alpha_L$ of the compute time divergence at fixed values of $K$.}
\newcolumntype{C}{>{\centering\arraybackslash}X}
\begin{tabularx}{\textwidth}{CCCC}
\toprule
\textbf{$K$} & \textbf{Network A}	& \textbf{Network B}	& \textbf{Network C}\\
\midrule
20  &  0.732(3) & 0.800(6) & 0.821(4) \\
40  & 0.752(2)  &0.820(4) & 0.872(4) \\
80  & 0.810(3) & 0.851(2) & 0.900(8) \\
160  & 0.810(4) & 0.800(2) & 0.924(4)\\
320    & 0.792(3) & 0.879(4) & 0.898(6)\\
\bottomrule
\end{tabularx}
\end{table}

                        The values of the exponents $\alpha_K(L)$ and $\alpha_L(K)$ weakly vary with the value of $L$ and $K$, respectively. The standard errors of $\alpha_K(L)$ and $\alpha_L(K)$ tend to remain consistent between smaller and larger ensemble sizes. The distribution of computed time, as depicted in the figure \ref{fig: Histo} for network A, does not follow a normal distribution. Consequently, increasing the ensemble size does not lead to a decrease in the standard error.
                        For each network, however, the value of $\alpha_K$ is significantly larger than $\alpha_L$. 
                        Thus, the expected compute time increases faster with $K$ than $L$. Given that larger values of $K$ and $L$ typically lead to a better result, i.e., a consensus partition with a larger $Q$, one might naively conclude that it is better to increase $L$ rather than $K$. But, to determine if that conclusion is, in fact, correct, the way that $Q$ increases with $K$ and $L$ must be taken into account.
                        
                        To do that, begin by noting that for any finite-size network, there is only a finite number of possible partitions. Many modularity-maximizing algorithms will consistently find the actual best partition for very small networks. As the network size and number of possible partitions grow, the task becomes harder; algorithms start to fail to find the exact solution and only provide estimates of the actual, or exact, best partition. RenEEL appears to do very well at finding the actual best partition of networks of size of up to a few thousand nodes~\cite{Guo2019}. Still, even it can only find estimates of the exact best partition of larger networks, such as the three we analyze in this paper.  As values of $K$ and $L$ increase, the estimates improve, and the value of $Q$ of the consensus partition approaches $Q_{\max}$, the Modularity of the exact best partition.
                        
                        \begin{figure}
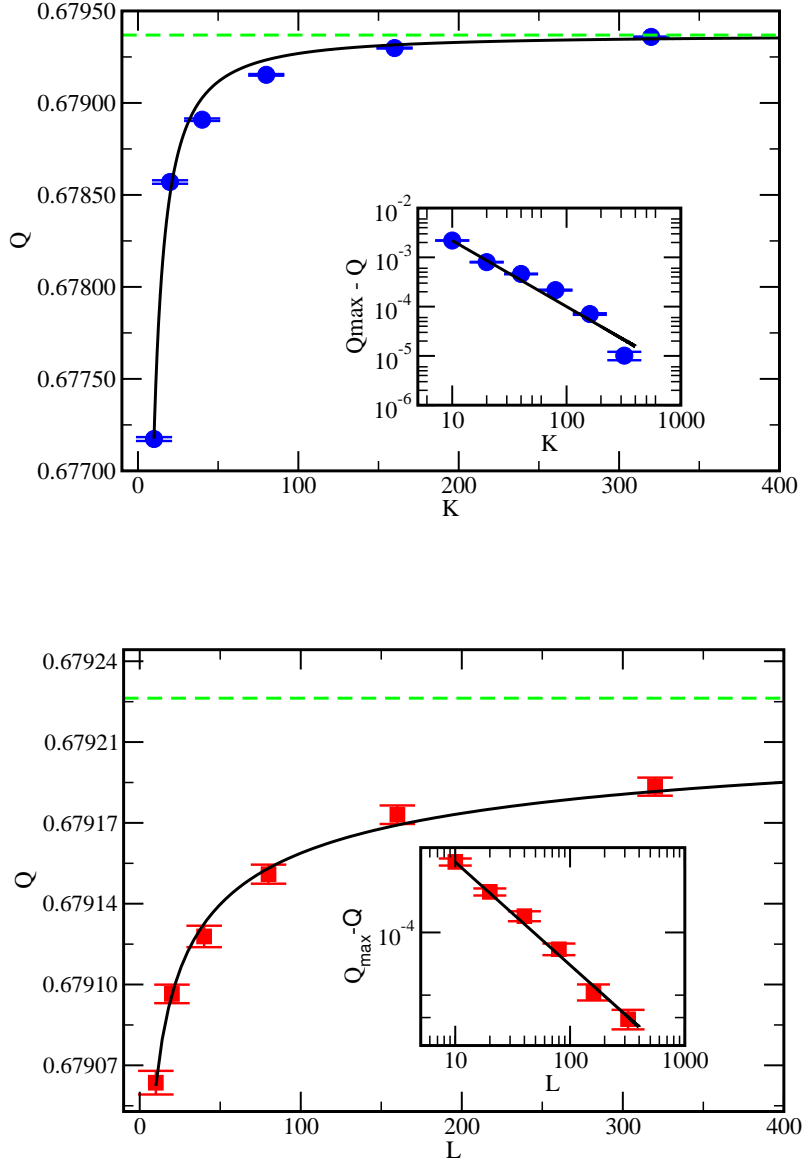

                            \centering
                            \begin{subfigure} 
                                \centering
                                \includegraphics[width=10.5 cm]{Mod_varyK_80.eps}
                            \end{subfigure}
                            
                            \vspace{1.5cm} 
                            \begin{subfigure}
                                \centering
                                \includegraphics[width=10.5 cm]{Mod_varyL_80.eps}
                            
                            \end{subfigure}
                            \caption{ Modularity $Q$ of RenEEL consensus partitions for Network A (a) as a function of $K$ for fixed $L=80$ (blue circles), and (b) as a function of $L$ for fixed $K=80$ (red squares). The black solid lines are fits to the data for a power-law approach to an estimated maximum value $Q_{\max}$ at the green dashed lines. The value of $Q_{\max}$ is 0.679368(6) in (a) and 0.679229(40) in (b). The insets show the data on log-log axes. The slope of the fit to the blue circles in the inset of (a) is $\gamma_K=1.34(2)$ and to the red squares in the inset of (b) is $\gamma_L=0.36(18)$.}
                            \label{fig:entirefigure}
                        \end{figure}
                        \unskip

                        To explore how the values of $Q$ of RenEEL's consensus partitions approach $Q_{\max}$
                        as a function of $K$ and $L$, the mean and standard error of $Q$ found in the  
                        runs that were made on each network were calculated as a function of $K$ and $L$.
                        The results are listed in the tables ~\ref{tab:Apen1a},~\ref{tab:Apen1b}, ~\ref{tab:Apen2a},~\ref{tab:Apen2b}, ~\ref{tab:Apen3a} and~\ref{tab:Apen3b} in the Appendix.
                        For a fixed value of $L$ or $K$, we find that 
                        $Q$ approaches a maximum value, $Q_{\max}$, as a power-law of the other ensemble size, 
                        \begin{eqnarray} 
                            Q & \approx & \left. Q_{max} - A_K K^{-\gamma_K}
                            \right|_{L\;{\rm fixed}} 
                            \nonumber
                            \\
                            \mbox{and} & ~ & \label{eq:n2}
                            \\
                            Q & \approx & \left. Q_{max} - A_L L^{-\gamma_L}
                            \right|_{K\;{\rm fixed}},
                            \nonumber
                        \end{eqnarray}
                        where the $A$'s are constants.
                        Figure~2 shows this behavior for Network A when $L$ and $K$ have fixed values of 80. The exact value of $Q_{\max}$ is unknown for Networks A, B, and C.  
                        Three-parameter, nonlinear least squares fits were used to determine the values of $Q_{\max}$, $A$, and $\gamma$. 
                        Tables \ref{tab:table3} and \ref{tab:table4} list the values of $Q_{\max}$ and $\gamma_K$, respectively, that result from fits at fixed values of $L$ for each of the three networks.
                        Similarly,
                        Tables \ref{tab:table5} and \ref{tab:table6} list the values of $Q_{\max}$ and $\gamma_L$ that result from fits at fixed values of $K$ for each of the three networks.
                        
                        \setcounter{table}{0}
                        \renewcommand{\thetable}{II\alph{table}}
                        \begin{table}
                        \captionsetup{justification=raggedright, singlelinecheck=false} 
                        \caption{\label{tab:table3} 
                        Values of asymptotic maximum Modularity $Q_{max}$ found at fixed values of $L$.}
                        \newcolumntype{C}{>{\centering\arraybackslash}X}
                        \begin{tabularx}{\textwidth}{CCCC}
                        \toprule
                        \textbf{$L$} & \textbf{Network A}	& \textbf{Network B}	& \textbf{Network C}\\
                        \midrule
                        10 & 0.679265(4) & 0.886761(2) & 0.742303(12) \\
                         20 & 0.679309(6) & 0.886784(6) & 0.742589(18)\\
                         40 & 0.679320(6) & 0.886814(2) & 0.742829(16)\\
                         80 & 0.679368(6) & 0.886825(4) & 0.742956(18)\\
                         160& 0.679370(8) & 0.886832(2) & 0.743013(16)\\
                         320 & 0.679377(4)& 0.886834(2) & 0.743168(20)\\
                        \bottomrule
                        \end{tabularx}
                        \end{table}
                        
                        \begin{table}
                        \caption{\label{tab:table4}
                        Values of scaling exponent $\gamma_K$ of the Modularity convergence to $Q_{\max}$ at fixed values of $L$.}
                        \newcolumntype{C}{>{\centering\arraybackslash}X}
                        \begin{tabularx}{\textwidth}{CCCC}
                        \toprule
                        \textbf{$L$} & \textbf{Network A}	& \textbf{Network B}	& \textbf{Network C}\\
                        \midrule
                        10 & 1.21(2)  & 2.16(4) & 1.18(18)\\
                         20 & 1.29(2)  & 2.15(6) & 1.31(20)\\
                         40 & 1.38(2)  & 2.01(8) & 1.28(20)\\
                         80 & 1.34(2)  & 2.04(4) & 1.71(16)\\
                         160& 1.35(4)  & 2.29(14) & 1.22(18)\\
                         320& 1.25(2)  & 2.05(6) & 1.29(20)\\
                        \bottomrule
                        \end{tabularx}
                        \end{table}

                        \setcounter{table}{0}
                        \renewcommand{\thetable}{III\alph{table}}
                        \begin{table}
                        \captionsetup{justification=raggedright, singlelinecheck=false} 
                        \caption{\label{tab:table5}
                        Values of asymptotic maximum Modularity $Q_{max}$ found at fixed values of $K$.}
                        \newcolumntype{C}{>{\centering\arraybackslash}X}
                        \begin{tabularx}{\textwidth}{CCCC}
                        \toprule
                        \textbf{$K$} & \textbf{Network A}	& \textbf{Network B}	& \textbf{Network C}\\
                        \midrule
                        10& 0.677480(34) & 0.886521(12)&    0.740259(170) \\
                        20& 0.678725(24)& 0.886763(10)&    0.741572(174)\\
                        40& 0.678963(36)& 0.886830(12)&    0.743070(172)\\
                        80& 0.679229(40)& 0.886832(10)&    0.743072(170)\\
                        160& 0.679246(26)& 0.886832(6)&   0.743077(172)\\
                        320& 0.679370(20)& 0.886833(8)&   0.743088(168)\\
                        \bottomrule
                        \end{tabularx}
                        \end{table}
                        
                        \begin{table}
                        \caption{\label{tab:table6}
                        Values of scaling exponent $\gamma_L$ of the Modularity convergence to $Q_{\max}$ at fixed values of $K$.}
                        \newcolumntype{C}{>{\centering\arraybackslash}X}
                        \begin{tabularx}{\textwidth}{CCCC}
                        \toprule
                        \textbf{$K$} & \textbf{Network A}	& \textbf{Network B}	& \textbf{Network C}\\
                        \midrule
                        10&  0.37(6) & 0.86(10)& 0.29(26) \\
                        20&  0.49(8) & 0.88(10)& 0.29(22)\\
                        40&  0.50(12) & 0.85(16)& 0.22(24)\\
                        80&  0.36(18) & 0.85(12)& 0.20(22) \\
                        160& 0.47(20) & 0.69(12)& 0.25(20) \\
                        320& 0.67(10) & 0.86(16)& 0.25(20) \\
                        \bottomrule
                        \end{tabularx}
                        \end{table}
                        
                        The fitted values of $Q_{\max}$ increase systematically with increasing $L$ and $K$ and converge to statistically equivalent values at the largest ensemble sizes studied (320) whether increasing $L$ or $K$. However, the values of $Q_{\max}$ are generally larger when fixing $L$ rather than $K$ when comparing results from when they are fixed at the same size. This fact implies that the maximum value of $Q$ is approached faster by fixing $L$ and increasing $K$ rather than doing the opposite. 
                        The rate of convergence to $Q_{\max}$ is quantified by the $\gamma$ exponents. 
                        The values of the exponents $\alpha_K(L)$ and $\alpha_L(K)$ depend on the network but only weakly vary with the value of $L$ and $K$, respectively. For each network, however, the value of $\alpha_K$ is significantly larger than $\alpha_L$. 
                        Thus, the expected compute time increases faster with $K$ than it does with $L$.
                        
                        \begin{figure}
                        \centering
                        \includegraphics[width = 10.5 cm]{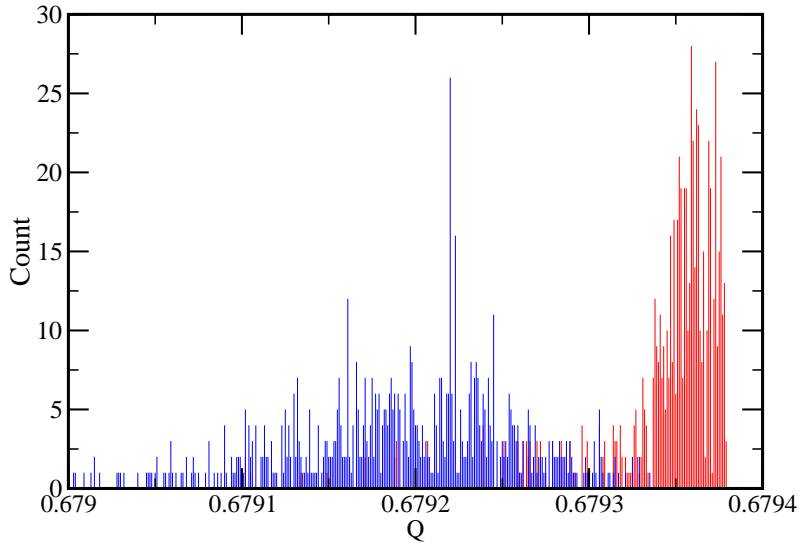}
                        
                        \caption{\label{fig: Histo} Distribution of $Q$ obtained for Network A. Blue bars correspond to the data for $L = 320$ and $K = 80$, and red bars to $L = 80$ and $K = 320$.}
                        \label{fig:p(q)}
                        \end{figure}
                        \unskip
                        To understand these results, recognize that finding the best partition is an extremal process that, when repeated, is akin to the process of record-breaking. Let's recall some of the theory of the extreme value statistics of record breaking~\cite{Schmittmann_1999}. Consider a sequence of independent and identically distributed random numbers 
                        \begin{equation}
                            \left\{ x_1, x_2, x_3, \ldots, x_t, \ldots \right\}
                        \end{equation}
                        chosen from a probability distribution of the form
                        \begin{equation}
                            p(x) = \mu B^{-\mu} (B-x)^{\mu-1}, \qquad 0\leq x \leq B,
                            \label{px}
                        \end{equation}
                        where $B$ is the maximum possible value of $x$, and define the \emph{record} $R(t)$ as the maximum value of $x$ in the first $t$ numbers in the sequence:
                        \begin{equation}
                            R(t)
                            \equiv
                              \max\left\{ x_1, x_2, x_3, \ldots, x_t\right\}.
                        \end{equation}
                        Then, in the limit of large $t$, the mean record will approach $B$ as 
                        \begin{equation}
                          \langle R(t) \rangle = B\left( 1  - \Gamma(1/\mu)\; t^{-1/\mu} \right),
                  \label{avgR}
                \end{equation}
            i.e., as a power-law function with exponent $1/\mu$.
            From this, we see that $\mu=1$ is a borderline case; from Eq.~\ref{px}, it is the case of a uniform distribution of $x$. If $\mu<1$, then $p(x)$ is maximal at $x=B$, and if $\mu>1$, then $p(x)$ vanishes as $x \rightarrow B$.

While the analogy with this simple, analytically tractable model of record-breaking is not perfect, Eqs.~\ref{eq:n2} can be compared with Eq.~\ref{avgR} by identifying $Q_{\max}$ with $B$ and $\gamma$ with $1/\mu$.
Then, the fact that empirically $\gamma_K>1$ and $\gamma_L<1$ suggests that the distributions of the $Q$ of the consensus partitions found by increasing $K$ and $L$ correspond to different cases. Namely, as $K$ is increased, the consensus partition $Q$ is likely to be near $Q_{\max}$ as it is for $\mu <1$, while as $L$ is increased, it is more likely to have a smaller value as it is for $\mu >1$. 
To confirm this, we made
800 runs of RenEEL analyzing Network A with $L=80$ and $K=320$ and with $L=320$ and $K=80$. 
Figure~\ref{fig: Histo} shows the consensus values of $Q$ found in those runs.
As expected, the values found with $L=80$ and $K=320$ (red bars) are much more likely to be near the maximum value than those found with $L=320$ and $K=80$ (blue bars).
\setcounter{table}{0}
\renewcommand{\thetable}{IV\alph{table}}
\begin{table}
\captionsetup{justification=raggedright, singlelinecheck=false} 
\caption{\label{tab:table7}
Values of scaling exponent $\frac{\gamma}{\alpha}$ at fixed values of $L$.}
\newcolumntype{C}{>{\centering\arraybackslash}X}
\begin{tabularx}{\textwidth}{CCCC}
\toprule
\textbf{$L$} & \textbf{Network A}	& \textbf{Network B}	& \textbf{Network C}\\
\midrule
20& 0.93(1)& 1.70(5)&    0.98(15)\\
40& 0.96(1)& 1.65(6)&    0.98(15)\\
80& 0.90(1)& 1.64(3)&    1.31(12)\\
160& 0.91(2)& 1.96(12)&  0.93(13)\\
320& 0.89(1)& 1.71(5)&   0.98(12)\\
\bottomrule
\end{tabularx}
\end{table}
\begin{table} 
\captionsetup{justification=raggedright, singlelinecheck=false} 
\caption{\label{tab:table8}
Values of scaling exponent $\frac{\gamma}{\alpha}$ at fixed values of $K$.}
\newcolumntype{C}{>{\centering\arraybackslash}X}
\begin{tabularx}{\textwidth}{CCCC}
\toprule
\textbf{$K$} & \textbf{Network A}	& \textbf{Network B}	& \textbf{Network C}\\
\midrule
20&  0.66(11) & 1.09(13)& 0.35(26)\\
40&  0.66(16) & 1.03(20)& 0.25(26)\\
80&  0.44(22) & 0.99(14)& 0.22(24) \\
160& 0.57(22) & 0.88(15)& 0.27(21) \\
320& 0.68(12) & 0.97(18)& 0.28(22) \\
\bottomrule
\end{tabularx}
\end{table}

We can now answer the central question of this paper: Given only limited computational resources, is it better to increase $K$ or $L$?
We have found that the average compute time grows faster with $K$ than with $L$, but also that the consensus $Q$ approaches $Q_{\max}$ faster with $K$ than with $L$. Does the consensus $Q$ approach $Q_{\max}$ as a function of compute time faster by increasing $K$ or $L$? To answer this invert Eqs.~\ref{eq2} and combine them with Eqs.~\ref{eq:n2} to obtain 
\begin{equation}
    Q_{max} - Q \sim \langle t \rangle^{-\gamma/\alpha}
\end{equation}
So, the larger $\gamma/\alpha$ is, the faster $Q$ approaches $Q_{max}$ as a function of average compute time.
Table~\ref{tab:table7} shows the values of $\gamma_K/\alpha_K$ at different fixed $L$ for the three networks.
Similarly, Tab.~\ref{tab:table8} shows the values of $\gamma_L/\alpha_L$ at different fixed $L$ for the three networks.
\newline
From these results, it can be clearly concluded that increasing $K$ rather than $L$, will cause 
$Q$ to approach $Q_{max}$ faster! With limited computational resources, is it better to increase $K$ rather than $L$.
\newline

\textbf{Author Contributions:} T.G., R.Z., and K.B. conceived of the presented idea. T.G. performed the numerical simulations and data analysis. K.B. verified the analytical methods and supervised the project. All authors discussed the results and contributed to the final manuscript. 
\newline

\textbf{Acknowledgments:} The authors would like to thank the Sabine cluster at the University of Houston, which was utilized for performing the simulations presented in this work.
\newline

\textbf{Conflicts of Interest:} The authors declare no conflict of interest. The funders had no role in the design of the study; in the collection, analyses, or interpretation of data; in the writing of the manuscript; or in the decision to publish the results.
\newline

\textbf{Data Availability:} The three networks used in this study are openly available at \href{https://sites.cc.gatech.edu/dimacs10/archive/clustering.shtml}{Data}. The raw data used for analysis are provided in the Appendix.
\newline

\textbf{Abbreviations:}
The following abbreviations are used in this manuscript:\\

\noindent 
\begin{tabular}{@{}ll}
RenEEL & Reduced Network Extremal Ensemble Learning\\
EEL & Extremal Ensemble Learning\\
\end{tabular}

\appendix  
\newcommand{\appendixsection}[1]{%
    \section*{Appendix #1} 
    \addcontentsline{toc}{section}{Appendix #1} 
}

\appendixsection{A}
\setcounter{table}{0}  
\renewcommand{\thetable}{A\arabic{table}}  
\begin{table}[h!]
\captionsetup{justification=raggedright, singlelinecheck=false} 
\caption{\label{tab:Apen1}Average computed time for Network A} 
\begin{tabular}{l c c c c c c}
\toprule 
& \multicolumn{5}{c}{\textbf{$K$}} \\ 
\cmidrule(l){2-7} 
\textbf{$L$} & 10 & 20 & 40 & 80 & 160 & 320\\ 
\midrule 
10 & 50.6(0.3) & 123(1) & 298(4) & 755(5) & 2065(57) & 5811(104)\\ 
20 & 59.7(0.3) &161.2(0.9) & 391(2) & 1094(6) & 2732(24) & 7241(42)\\ 
40 & 72.0(0.3) & 202(1) & 564(3) & 1617(9) & 3972(21) & 10839(58)\\ 
80 & 93(1) & 326(2) & 960(5) & 2427(13) & 6857(41) & 18568(212)\\ 

160 & 145.1(0.9) & 501(3) & 1559(10) & 4641(29) & 12123(109) & 34007(245)\\ 
320 & 215(1)& 919(5) &2908(16) & 8385(56) & 22964(220) & 58568(388)\\
\midrule 
\midrule 
\end{tabular}

\label{tab:time1} 
\end{table}

\begin{table}[h] 
\captionsetup{justification=raggedright, singlelinecheck=false} 
\caption{\label{tab:Apen1a}Modularity Value for Network A} 
\begin{tabular}{l c c c } 
\toprule 
& \multicolumn{3}{c}{\textbf{$K$}} \\ 
\cmidrule(l){2-4} 
\textbf{$L$} & 10 & 20 & 40 \\ 
\midrule 
10 & 0.67691112(300) & 0.67829673(134) & 0.67877639(779) \\ 
20 & 0.677076101(1973) & 0.67844664(446) & 0.67881534(751)\\ 
40 & 0.67715786(897) & 0.67853550(1947) & 0.67887617(823) \\ 
80 & 0.67717013(1549) & 0.67851858(941) & 0.67890690(564) \\ 
160 & 0.67730312(899) & 0.67863411(271) & 0.67890899(716) \\ 
320 & 0.67732834(2819) & 0.67865861(980) & 0.67893038(621) \\
\midrule 
\midrule 
\end{tabular}

\label{tab:time1} 
\end{table}
\begin{table}[h] 
\captionsetup{justification=raggedright, singlelinecheck=false} 
\caption{\label{tab:Apen1b}Modularity Value for Network A} 
\begin{tabular}{l c c c } 
\toprule 
& \multicolumn{3}{c}{\textbf{$K$}} \\ 
\cmidrule(l){2-4} 
\textbf{$L$} & 80 & 160 & 320\\ 
\midrule 
10 &  0.67906248(512) &  0.67921816(584) & 0.67928542(149)\\ 
20 & 0.67910093(401) & 0.67925767(629) & 0.67932669(188)\\ 
40 & 0.67912582(465) & 0.67928322(294) & 0.67933913(177)\\ 
80 & 0.67915278(412) & 0.67929836(261) & 0.67935888(196)\\ 
160 & 0.67917852(401) & 0.67931103(397) & 0.67936292(202)\\ 
320 & 0.67919068(391) & 0.67932548(421) & 0.67936962(97)\\
\midrule 
\midrule 
\end{tabular}

\label{tab:time1} 
\end{table}
\newpage
\begin{table}[h] 
\captionsetup{justification=raggedright, singlelinecheck=false} 
\caption{\label{tab:Apen2}Average computed time for Network B} 
\begin{tabular}{l c c c c c c} 
\toprule 
& \multicolumn{5}{c}{\textbf{$K$}} \\ 
\cmidrule(l){2-7} 
\textbf{$L$} & 10 & 20 & 40 & 80 & 160 & 320\\ 
\midrule 
10 & 13.84(0.07) & 30.7(0.2) & 63.1(0.1) & 143.1(0.6) & 293(1) & 625(3)\\ 
20 & 16.37(0.09) & 39.7(0.2) & 88.5(0.3) & 196(1) & 426(2) & 833(4)\\ 
40 & 20.2(0.1) & 60.5(0.3)& 130(1) & 293(1) & 649(3) & 1340(7)\\ 
80 & 31.1(.2)& 102(0.9) & 256(2) & 514(3) & 1199(8) & 2604(22)\\ 
160 & 46.8(0.3) & 170(1) & 431(3) & 969(6) & 2036(13) & 4363(26)\\ 
320 & 73.5(0.6) & 282(2) & 877(6) & 1955(14) & 4016(29) & 8214(51)\\
\midrule 
\midrule 
\end{tabular}

\label{tab:time1} 
\end{table}

\vspace{7mm}

\begin{table}[h!] 
\captionsetup{justification=raggedright, singlelinecheck=false} 
\caption{\label{tab:Apen2a}Modularity Value for Network B}
\begin{tabular}{l c c c} 
\toprule 
& \multicolumn{3}{c}{\textbf{$K$}} \\ 
\cmidrule(l){2-4} 
\textbf{$L$} & 10 & 20 & 40\\ 
\midrule 
10 & 0.88642816(536) & 0.88674884(440) & 0.88671202(790)\\ 
20 & 0.88647127(1895) & 0.88674307(594) & 0.88675001(918) \\ 
40 & 0.88641007(295) & 0.88677498(371) & 0.8867800(1021) \\ 
80 & 0.88649938(1104) & 0.88674605(430) & 0.88678501(865) \\ 
160 & 0.88655162(1003) & 0.88676799(246) & 0.88680202(672)\\ 
320 & 0.88655147(675) & 0.88676975(315) & 0.88681905(495)\\
\midrule 
\midrule 
\end{tabular}
\label{tab:time1} 
\end{table}
\vspace{5mm}
\begin{table}[h!]
\captionsetup{justification=raggedright, singlelinecheck=false} 
\caption{\label{tab:Apen2b}Modularity Value for Network B} 
\begin{tabular}{l c c c} 
\toprule 
& \multicolumn{3}{c}{\textbf{$K$}} \\ 
\cmidrule(l){2-4} 
\textbf{$L$} & 80 & 160 & 320\\ 
\midrule 
10 & 0.88673690(1014) &  0.88674209(745) & 0.88675501(832)\\ 
20 & 0.88677493(799) & 0.88678016(634) & 0.88678290(405)\\ 
40 & 0.88680491(805) & 0.88680996(678) & 0.886801492(418)\\ 
80 & 0.88680099(697) & 0.88681788(395) & 0.88681484(302)\\ 
160 & 0.88682706(707) & 0.88683401(471) & 0.88683493 (247)\\ 
320 & 0.88683201(381) & 0.88683687(382) & 0.88683980(201)\\
\midrule 
\midrule 
\end{tabular}

\label{tab:time1} 
\end{table}
\vspace{5mm}
\begin{table}
\captionsetup{justification=raggedright, singlelinecheck=false} 
\caption{\label{tab:Apen3}Average computed time for Network C} 
\begin{tabular}{l c c c c c c} 
\toprule 
& \multicolumn{5}{c}{\textbf{$K$}} \\ 
\cmidrule(l){2-7} 
\textbf{$L$} & 10 & 20 & 40 & 80 & 160 & 320\\ 
\midrule 
10 &  139.5(0.7) & 295(2) & 640(5) & 1466(10)& 3483(23) & 8485(70))\\ 
20 & 181(1) & 415(2) & 1010(9) & 2389(16) & 5806(35)& 14274(108)\\ 
40 & 295(2) & 716(6) & 1754(10) & 4350(31) & 10846(82) & 25722(161)\\ 
80 & 484(4) & 1251(13) & 3203(30) & 8033(68) & 20680(162) & 45012(402)\\ 
160 & 909(9) & 2340(28) & 6096(52) & 15600(123) & 41308(344) & 87912(801)\\ 
320 & 1790(19)& 4600(53) & 11936(106) & 29605(249) & 77968(620) & 170610(1001)\\
\midrule 
\midrule 
\end{tabular}
\label{tab:time1} 
\end{table}
\begin{table}
\captionsetup{justification=raggedright, singlelinecheck=false} 
\caption{Modularity Value for Network C} 
\label{tab:Apen3a}
\begin{tabular}{l c c c} 
\toprule 
& \multicolumn{3}{c}{\textbf{$K$}} \\ 
\cmidrule(l){2-4} 
\textbf{$L$} & 10 & 20 & 40 \\ 
\midrule 
10 & 0.73982575(12141) & 0.74042051(10865) & 0.74119763(9309) \\ 
20 & 0.73946114(16171) & 0.74082625(12067) & 0.74199647(6703) \\ 
40 & 0.73847614(7023) & 0.74163987(6899) & 0.74230803(7021) \\ 
80 & 0.73959559(15273) & 0.74192572(9061) & 0.74221369(6149) \\ 
160 & 0.73966281(10312) & 0.7411455(13235) & 0.7424959(6301) \\ 
 320 & 0.74011840(13382) & 0.74154016(12306) & 0.74209565(8293) \\
\midrule 
\midrule 
\end{tabular}
\end{table}

\begin{table} 
\captionsetup{justification=raggedright, singlelinecheck=false}  
\caption{Modularity Value for Network C} 
\label{tab:Apen3b}
\begin{tabular}{l c c c} 
\toprule 
& \multicolumn{3}{c}{\textbf{$K$}} \\ 
\cmidrule(l){2-4} 
\textbf{$L$} & 80 & 160 & 320\\ 
\midrule 
10 & 0.74155587(8022) &  0.74188600(8036) & 0.74189623(8136)\\ 
20 &  0.74220959(7080) & 0.74202369(8062) & 0.74203165(8518)\\ 
40 & 0.74230761(6930) & 0.74229803(8301) & 0.74216325(8772)\\ 
80 & 0.74239582(7500) & 0.74251785(6805) & 0.74251525(6181)\\ 
160 & 0.7425311(7104) & 0.7426539(7409) & 0.7426924(6417)\\ 
320 & 0.74273495(6921) & 0.74270134(4921) & 0.74271412(2012)\\
\midrule 
\midrule 
\end{tabular}
\end{table}

\appendixsection{B}

\begin{figure}[h!]
\centering
\includegraphics[width = 10.2 cm]{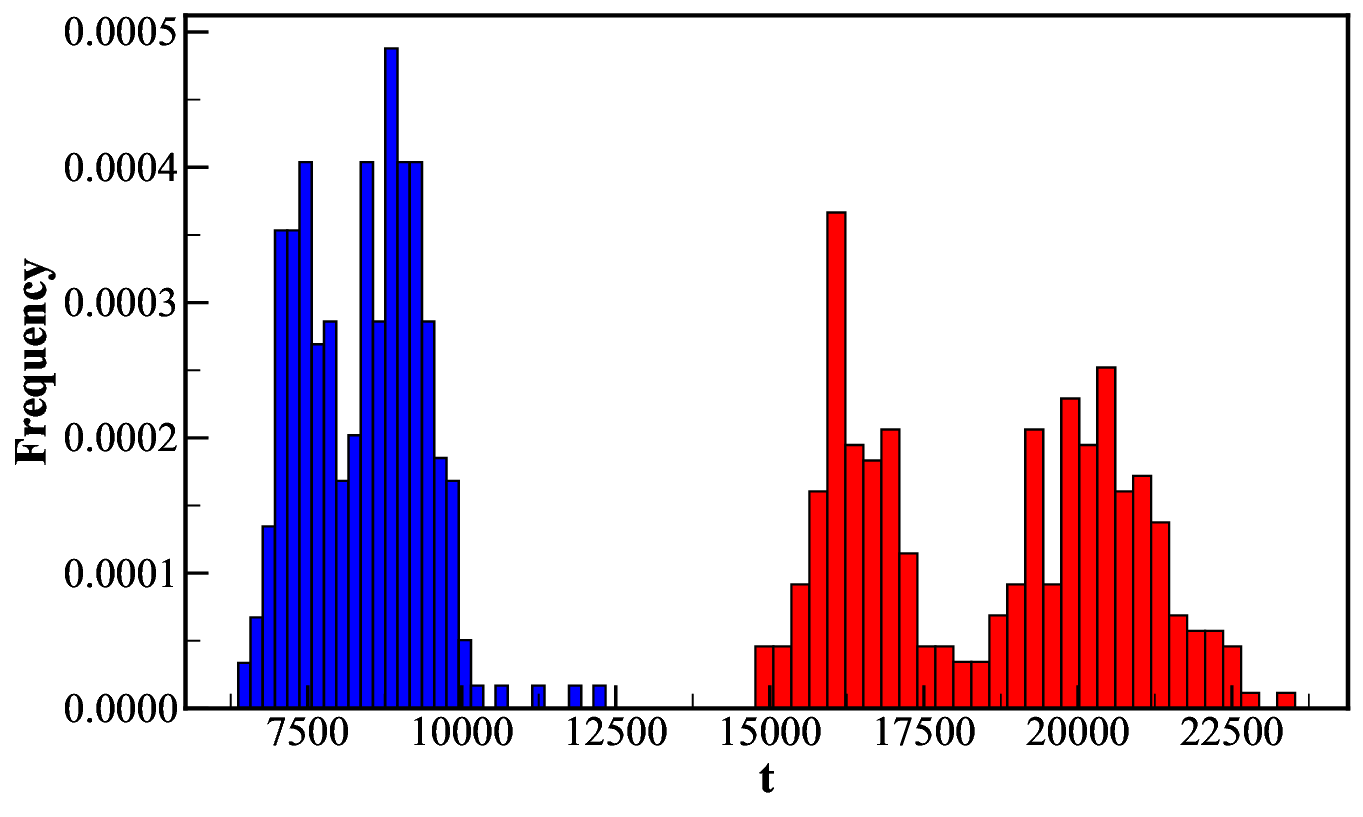}

\caption{\label{fig: Histo} Distribution of required computed time to find the consensus partition for Network A. Blue bars correspond to the data for $L = 320$ and $K = 80$, and red bars to $L = 80$ and $K = 320$.}
\label{fig:p(q)}
\end{figure}

\newpage
\bibliographystyle{unsrt}  
\bibliography{main}      

\providecommand{\noopsort}[1]{}\providecommand{\singleletter}[1]{#1}%
\begin{thebibliography}{10}

\bibitem{Fortunato_2010}
Santo Fortunato.
\newblock Community detection in graphs.
\newblock {\em Physics Reports}, 486(3-5):75--174, feb 2010.

\bibitem{Newman_2004_A}
M.~E.~J. Newman and M.~Girvan.
\newblock Finding and evaluating community structure in networks.
\newblock {\em Phys. Rev. E}, 69:026113, Feb 2004.

\bibitem{Schaub_2017}
Michael~T. Schaub, Jean-Charles Delvenne, Martin Rosvall, and Renaud Lambiotte.
\newblock The many facets of community detection in complex networks.
\newblock {\em Applied Network Science}, 2(1):4, 2017.

\bibitem{Peel_2017}
Leto Peel, Daniel~B. Larremore, and Aaron Clauset.
\newblock The ground truth about metadata and community detection in networks.
\newblock {\em Science Advances}, 3(5):e1602548, 2017.

\bibitem{Wolpert_2005}
D.H. Wolpert and W.G. Macready.
\newblock Coevolutionary free lunches.
\newblock {\em IEEE Transactions on Evolutionary Computation}, 9(6):721--735, 2005.

\bibitem{Newman_2006_B}
M.~E.~J. Newman.
\newblock Modularity and community structure in networks.
\newblock {\em Proceedings of the National Academy of Sciences}, 103(23):8577--8582, 2006.

\bibitem{KuninJN_2023}
Alexander~B. Kunin, Jiahao Guo, Kevin~E. Bassler, Xaq Pitkow, and Kre{\v s}imir Josi{\'c}.
\newblock Hierarchical modular structure of the drosophila connectome.
\newblock 2023.

\bibitem{ZAMANIESFAHLANI2021}
Farnaz {Zamani Esfahlani}, Youngheun Jo, Maria~Grazia Puxeddu, Haily Merritt, Jacob~C. Tanner, Sarah Greenwell, Riya Patel, Joshua Faskowitz, and Richard~F. Betzel.
\newblock Modularity maximization as a flexible and generic framework for brain network exploratory analysis.
\newblock {\em NeuroImage}, 244:118607, 2021.

\bibitem{Brandes_2008}
Ulrik Brandes, Daniel Delling, Marco Gaertler, Robert Gorke, Martin Hoefer, Zoran Nikoloski, and Dorothea Wagner.
\newblock On modularity clustering.
\newblock {\em IEEE Transactions on Knowledge and Data Engineering}, 20(2):172--188, 2008.

\bibitem{Brandes_2006}
Ulrik Brandes, Daniel Delling, Marco Gaertler, Robert G{\"{o}}rke, Martin Hoefer, Zoran Nikolski, and Dorothea Wagner.
\newblock On modularity - np-completeness and beyond.
\newblock Technical Report~19, {Universit{\"{a}}t Karlsruhe (TH)}, 2006.

\bibitem{Blondel_2008}
Vincent~D Blondel, Jean-Loup Guillaume, Renaud Lambiotte, and Etienne Lefebvre.
\newblock Fast unfolding of communities in large networks.
\newblock {\em Journal of Statistical Mechanics: Theory and Experiment}, 2008(10):P10008, 2008.

\bibitem{Ovelgnne_2010}
Michael Ovelg{\"o}nne and Andreas Geyer-Schulz.
\newblock Cluster cores and modularity maximization.
\newblock {\em 2010 IEEE International Conference on Data Mining Workshops}, pages 1204--1213, 2010.

\bibitem{Sun2009}
Y.~Sun, B.~Danila, K.~Josić, and K.~E. Bassler.
\newblock Improved community structure detection using a modified fine-tuning strategy.
\newblock {\em Europhysics Letters}, 86(2):28004, may 2009.

\bibitem{Trevino_2015}
Santiago Treviño, Amy Nyberg, Charo I~Del Genio, and Kevin~E Bassler.
\newblock Fast and accurate determination of modularity and its effect size.
\newblock {\em Journal of Statistical Mechanics: Theory and Experiment}, 2015(2):P02003, feb 2015.

\bibitem{Liu_2019}
Zhiyuan Liu and Yinghong Ma.
\newblock A divide and agglomerate algorithm for community detection in social networks.
\newblock {\em Information Sciences}, 482:321--333, 2019.

\bibitem{Guo2019}
Jiahao Guo, Pramesh Singh, and Kevin~E. Bassler.
\newblock Reduced network extremal ensemble learning scheme for community detection in complex networks.
\newblock {\em Scientific Reports}, 9(1):14234, 2019.

\bibitem{Newman_2011}
Mark Newman.
\newblock Network data.
\newblock \url{http://www-personal.umich.edu/~mejn/netdata/}, Accessed: 2023.

\bibitem{Newman_2001}
M.~E.~J. Newman.
\newblock The structure of scientific collaboration networks.
\newblock {\em Proceedings of the National Academy of Sciences}, 98(2):404--409, 2001.

\bibitem{Schmittmann_1999}
B.~Schmittmann and R.~K.~P. Zia.
\newblock {\textquotedblleft}weather{\textquotedblright} records: Musings on cold days after a long hot indian summer.
\newblock {\em American Journal of Physics}, 67(12):1269--1276, dec 1999.

\end{thebibliography}

\end{document}